\documentclass[aps,prl,twocolumn,amsmath,amssymb,showpacs,superscriptaddress,floatfix]{revtex4-1}
\usepackage{graphicx}
\usepackage{dcolumn}
\usepackage{bm}
\usepackage{amsfonts}
\usepackage{amsmath}
\usepackage{ulem}
\usepackage{color}
\usepackage{hyperref}
\newcommand{\fab}[1]{}

\begin{document}

\title{Signature of interaction in dc transport of ac gated Quantum Spin Hall edge states} 

\author{Fabrizio Dolcini}
\email{fabrizio.dolcini@polito.it}

\affiliation{Dipartimento di Fisica del Politecnico di Torino, I-10129 Torino, Italy}
\begin{abstract}
In the presence of a scattering potential, electron transport in a quantum wire is known to be dramatically modified by backward scattering and unaffected by forward scattering processes. We show that the scenario is quite different in Quantum Spin Hall effect edge states coupled at a constriction. The helical nature of these states leads to the appearance of a forward scattering spin channel that is absent in other Luttinger liquid realizations. Suitably applied ac gate voltages can thus operate on the spin of  electrons tunneling  across the constriction, and induce in the dc tunneling current a cusp pattern  that represents the signature of the edge state  electronic interaction.
\end{abstract}

\pacs{71.10.Pm, 73.23.-b, 85.75.-d}
\maketitle


The nanotechnological advances of last two decades have allowed to achieve various solid state realizations of  one-dimensional (1D) electronic systems, such as semiconductor  wires~\cite{QW}, carbon nanotubes~\cite{SWNT}, and Quantum Hall edge states\cite{QHE}, where electron-electron correlations  lead to a Luttinger liquid~(LL) behavior~\cite{LL}.
One of the most striking features that distinguishes a  LL  from  a non-interacting system is its unconventional electron transport. This can already be seen from  the simple case of the  current flowing through an interacting wire in the presence of one single localized impurity, modeled as a delta-like scattering potential $\lambda \, \delta(x)$. Indeed, a well known result of LL theory is that, no matter how weak the scatterer strength~$\lambda$ is, the current-voltage characteristics exhibits a zero-bias anomaly with an interaction dependent power-law   at low voltage bias~\cite{kane-fisher,glazman}. Such peculiar behavior  originates from the interplay between electron-electron interaction and single-particle backward scattering (BS) processes occurring at the impurity. In general, besides BS, an impurity   also gives rise to   forward  scattering (FS) processes, which are known to have no effect on the current of a quantum wire, though (see Ref.~\cite{vondelft}). Thus, while BS has been widely discussed~\cite{LL,kane-fisher,glazman,vondelft}, impurity FS terms are often omitted  in models of quantum wires, with the underlying understanding that such terms never affect   transport. 

In this paper we show that the situation can be quite different in the recently discovered edge states of Quantum Spin Hall effect (QSHE) systems~\cite{QSHE}. These 1D electronic states, flowing at the edges of HgTe/CdTe quantum wells, are characterized by a tight connection between the direction of motion and spin orientation, and  represent a new type of LLs, called {\it helical} Luttinger liquids~\cite{hLL}. Here we shall show that in such systems a different type of FS emerges, which is absent in other 1D realizations, and which strongly affects transport via the spin channel. In particular we shall demonstrate that such effect can lead to a cusp pattern in the current-voltage characteristics that represents the signature of electron-electron interaction in the QSHE edge states. \\
\begin{figure}[h]
\centering
\includegraphics[width=0.8\linewidth]{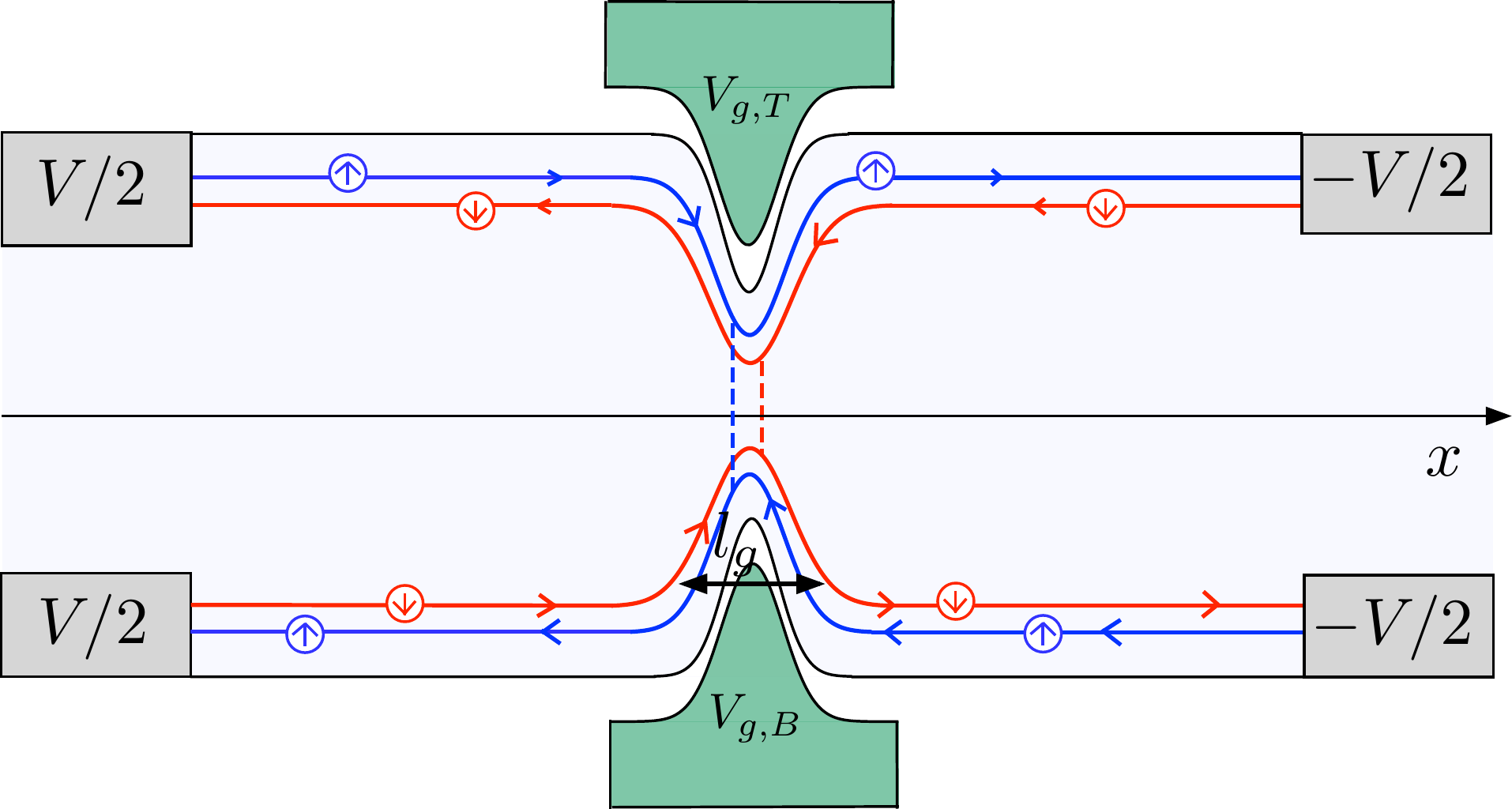} 
\caption{\label{fig-setup} (color on-line) Helical edge states flow in a QSHE bar (circled arrows denote spin orientation). Two ac gate voltages $V_{g,T}$ and $V_{g,B}$ applied with opposite phase across a QPC affect the spin of the tunneling electrons and, in the presence of electron-electron interaction, modify the dc current. }
\end{figure}

In order to illustrate this phenomenon, we first briefly  recall  the usual case of a quantum wire. We denote by   $\mathcal{H}=\mathcal{H}_0+\mathcal{H}_{imp}^{FS}+\mathcal{H}_{imp}^{BS}$   the Hamiltonian of a  quantum wire with an impurity, where $\mathcal{H}_0$ is the LL Hamiltonian of the clean interacting wire,  
$\mathcal{H}^{BS}_{imp}= \lambda \, \sum_{\sigma}   (\Psi^\dagger_{R \sigma} \Psi^{}_{L \sigma} +\Psi^\dagger_{L \sigma} \Psi^{}_{R \sigma}) \, \,  |_{x=0}$ describes the impurity BS term, and $\mathcal{H}^{FS}_{imp}= \lambda \, \sum_{\sigma}   (\Psi^\dagger_{R \sigma}  \Psi^{}_{R \sigma}+\Psi^\dagger_{L \sigma}  \Psi^{}_{L \sigma}) \, |_{x=0} \, \,$
the impurity FS term. Here $\Psi_{r \sigma}$ ($r=R/L=\pm$) denotes the right(left) moving component  of the electron field operator with  the  spin orientation $\sigma =\uparrow,\downarrow$. A handwaving argument to see that impurity FS has no effect on transport is to observe that, since  $\mathcal{H}^{FS}_{imp}$ couples to the sum of right and left movers, it cannot affect their difference, i.e. the current. At a more formal level one  expresses the electron fields $\Psi_{r \sigma}$   using the Bosonization identity~\cite{LL}, i.e. $\Psi_{r \sigma}(x)=  (2 \pi a)^{-1/2} \kappa_{r \sigma} \, \exp{[  i r \sqrt{\frac{\pi}{2}} (\Phi_{c}+\sigma \Phi_{s}+r (\Theta_{c}+\sigma \Theta_{s}))(x)]}$,
where $a$ denotes the underlying lattice spacing and $\kappa_{r \sigma}$ the Klein factors ensuring the anticommutation of different  species. The bosonic fields  $\Phi_{c(s)}(x)$  describe the charge and spin degrees of freedom, respectively, and $\Theta_{c(s)}(x)$ their dual fields,  fulfilling  the  commutation relations $[ \Phi_{\nu}(x), \Theta_{\nu^\prime}(y)]= i   \,  \delta_{\nu, \nu^\prime} \, \mbox{sgn}(x-y)/2$, with $\nu=c,s$.    Introducing now non-local fields $\xi_{ \nu \pm}(x) \doteq (\Phi_{\nu}(x) +\Theta_\nu(x)\pm (\Phi_{\nu}(-x)-\Theta_\nu(-x)))/2$, it is possible to show~\cite{vondelft}  that
the impurity terms   depend on $\xi_{\nu \pm}$ in the following way 
\begin{eqnarray}
\mathcal{H}^{BS}_{imp}& =& \mathcal{H}^{BS}_{imp}(\xi_{c+},\xi_{s+}) \label{BS-bos1} \\
\mathcal{H}^{FS}_{imp}& =& \mathcal{H}^{FS}_{imp}(\xi_{c-}) \quad, \label{FS-bos1}  
\end{eqnarray}
whereas the term $\mathcal{H}_0$ splits into a sum 
$\mathcal{H}_0 = \sum_{\nu=c,s} \sum_{\alpha=\pm} \mathcal{H}_{0 \nu \alpha}(\xi_{\nu \alpha})$ of independent terms ($[\xi_{\nu +}(x), \xi_{\nu^\prime -}(y)]=0$).
Since the term $\mathcal{H}^{FS}_{imp}$ is linear in $\xi_{c -}$, and the term $\mathcal{H}_{0 c -}$   is quadratic in $\xi_{c-}$,   the former term can be gauged away  simply by a unitary transformation~\cite{vondelft}.  Importantly, in doing that, a crucial point is that  $\mathcal{H}^{FS}_{imp}$ and $\mathcal{H}^{BS}_{imp}$  depend on {\it different} fields, as shown in Eqs.(\ref{BS-bos1})-(\ref{FS-bos1}), so that in a quantum wire FS and BS terms are  independent. Notice that such argument also holds for a time-dependent impurity, so that previous studies analyzing such situation have focussed on   time-dependent BS terms~\cite{feldman-gefen,chamon,komnik}. \\

Let us now discuss the case of the edge states of a QSHE system.     The helical properties  imply that, at each boundary of a HgTe/CdTe quantum well, a Kramers  pair  of counter-propagating  states appears, so that at the -say- top boundary right- (left-) moving electrons are characterized by spin-$\uparrow$ (spin-$\downarrow$) only. The opposite occurs at the bottom  boundary,  as  shown in Fig.\ref{fig-setup}, and each boundary carries both  charge and spin. Including  both intra- and inter-edge electron-electron interaction,  the system of the two boundaries can be described by a LL Hamiltonian~\cite{hLL}
\begin{equation}
\mathcal{H}_0 = \frac{\hbar}{2} \sum_{\nu=c,s} \int dx \left[ v_\nu K_\nu (\partial_x \Theta_\nu)^2+ \frac{v_\nu}{K_\nu} (\partial_x \Phi_\nu)^2 \right]
\end{equation}
where $v_c=v_s=v$ and $K_c=K_s^{-1}=K$ are   interaction dependent parameters, with $v K^{\pm 1}=v_F (1+(2U_1\pm U_2)/2 \pi \hbar v_F)$. Here $v_F$ denotes the Fermi velocity, and $U_1$ and $U_2$  the intra- and inter-edge interaction.  The sample is assumed infinitely long in the longitudinal direction~$x$.   The peculiarity of helical nature is encoded in the fact that,  differently from   quantum wires where the spin channel is essentially non interacting ($K_s =1$) due to SU(2) symmetry, in the QSHE edge states the spin channel is  characterized by an effective attractive interaction $K_s >1$~\cite{hLL,liu}.

Scattering from impurities along one boundary is prevented from topological protection. Nevertheless, by etching the quantum well over a short region to form a Quantum Point Contact (QPC), the two boundaries are brought close to each other  [see Fig.\ref{fig-setup}], inducing inter-boundary tunneling~\cite{hLL,liu,mio}. Due to the helical nature of the edges, an electron flowing along   a given direction can only tunnel to the other boundary  by reversing its group velocity~\cite{nota}. Tunneling thus plays the same role as the BS in a  quantum wire, and is described by a term
\begin{eqnarray}
\mathcal{H}^{}_{tun}  =  \displaystyle   \left. \frac{\hbar v_{\rm F} \gamma}{2\pi a} \!  \! \! \sum_{m=\pm, \sigma=\uparrow, \downarrow} \! \!  \! \! \! \! m   \,   \kappa_{L \sigma} \kappa_{R \sigma}  \,  \, e^{i m   \sqrt{2 \pi} (\xi_{c+} + \sigma \xi_{s+})}\right|_{x=0}  \label{Htun-p} 
\end{eqnarray}
where $\gamma$ is the dimensionless tunneling amplitude. Indeed Eq.(\ref{Htun-p}) has the same form as (\ref{BS-bos1}), and we shall utilize the expression `backward scattering' (`BS') to emphasize such analogy.
Furthermore, with the voltages $V_{g,T}$ and $V_{g,B}$ of two gates located at the two sides of the constriction, and coupled to the edge states over a lengthscale~$l_g$ [see Fig.\ref{fig-setup}],
one  generates local FS potential terms,
 namely $V_{g,T} (\rho_{R\uparrow}+\rho_{L\downarrow})$ for the top boundary, and $V_{g,B} (\rho_{R\downarrow}+\rho_{L\uparrow})$ for the bottom boundary,  
where $\rho_{r \sigma}=e \, \Psi^\dagger_{r \sigma} \Psi^{}_{r \sigma}$.

Despite these similarities,   important differences emerge with respect to the case of a quantum wire. 
Indeed,  due to the space separation between the edge states, the two  gates voltages couple differently to the edges, so that here two independently tunable FS terms appear. This difference  can be highlighted by making the   (unnecessary for the final result) assumption that  the electron fields are not significantly varying   along~$l_g$. Then, utilizing the non-local charge and spin fields $\xi_{\nu\pm}$, one obtains
\begin{eqnarray}
\mathcal{H}^{}_{gate}  &=& \sqrt{\frac{2}{\pi}} \left. e\, l_g \left( V_{g,c}  \, \partial_x \xi_{c-}+ V_{g,s}  \,  \partial_x\xi_{s+}   \right) \right|_{x=0} \label{Hgate} 
\end{eqnarray}
where $V_{g,c/s}=(V_{g,T}\pm V_{g,B})/2$. The first term of the r.h.s. of  Eq.(\ref{Hgate}), controlled by $V_{g,c}$, is the usual charge density coupling  also present in the term (\ref{FS-bos1}) of an impurity  in a quantum wire. As observed above, it has no effect on dc transport. However, Eq.(\ref{Hgate}) exhibits an additional FS channel, which represents a coupling to the spin current and is controlled by spin gate $V_{g,s}$. Notably, such  field $\xi_{s+}$ appears also in the `BS' term (\ref{Htun-p}), so that  for the spin channel FS and BS processes are not independent, and the former cannot be simply gauged away.  Thus, although the FS term {\it alone} cannot induce a dc current directly, in QSHE edge states it can in principle operate indirectly, by affecting the BS term. Indeed, generalizing now (\ref{Hgate}) to arbitrary space and time-dependent profile $V_{g,s}=V_{g,s}(x,t)$ around the constriction,  one can show that the additional spin FS term   leads to a shift $\xi_{s+}\rightarrow \xi_{s+}+\xi_{s+}^0$ in the exponent of tunneling term~(\ref{Htun-p}), where
\begin{eqnarray}
\xi_{s+}^0(t) &=& -\frac{1}{\sqrt{2 \pi}} \frac{e}{\hbar v} \int V_{g,s}(x^\prime,t-\frac{|x^\prime|}{v}) \,dx^\prime  \quad. \label{xis+0}
\end{eqnarray}
This represents a $V_{g,s}$-dependent renormalization of the phase of the tunneling amplitude $\gamma$. Since  only  phase differences at the tunneling point matter, the spin FS term can affect dc transport in the presence of a time-dependent spin gate voltage $V_{g,s}(x^\prime,t)$.\\

Thus, the  above outlined difference between the two terms of Eq.(\ref{Hgate})  also finds, mutatis mutandis, an interpretation in the context of photon-assisted tunneling~\cite{aguado}, where one studies the effect of an ac gate voltage on the dc current. For an energy independent scatterer (as a single impurity or tunneling term is) an ac gate voltage is known to yield no dc effect~\cite{butti}. This holds for the conventional coupling to the charge degree of freedom. However,  an electron tunneling across the QPC transfers not only charge but also spin. Helical QSHE edge states offer the possibility to `photon-assist' tunneling via the spin-channel, through the second term on the r.h.s. of  Eq.(\ref{Hgate}). 
It is worth noticing, at this point, that also in QHE systems the edge states are space separated~\cite{QHE} and, when a gate voltage difference $V_{g,T}-V_{g,B}$ is applied across a QPC  in a Hall bar, an additional FS term arises  w.r.t.  quantum wires. However, since QHE the edge states are chiral, such FS term is the charge current. It thus breaks time-reversal symmetry (TRS) and still involves the usual   degree of freedom, the charge, which exhibits  repulsive interaction ($K_c \le 1$). In contrast, since in QSHE the edge states are helical, the additional coupling preserves TRS and involves the spin channel, which is characterized by an attractive interaction $K_s=K^{-1} \ge 1$. In this respect, the QSHE edge states exhibit an unconventional effect that cannot be addressed in other systems.

\begin{figure}[h]
\centering
\includegraphics[width=0.9\linewidth]{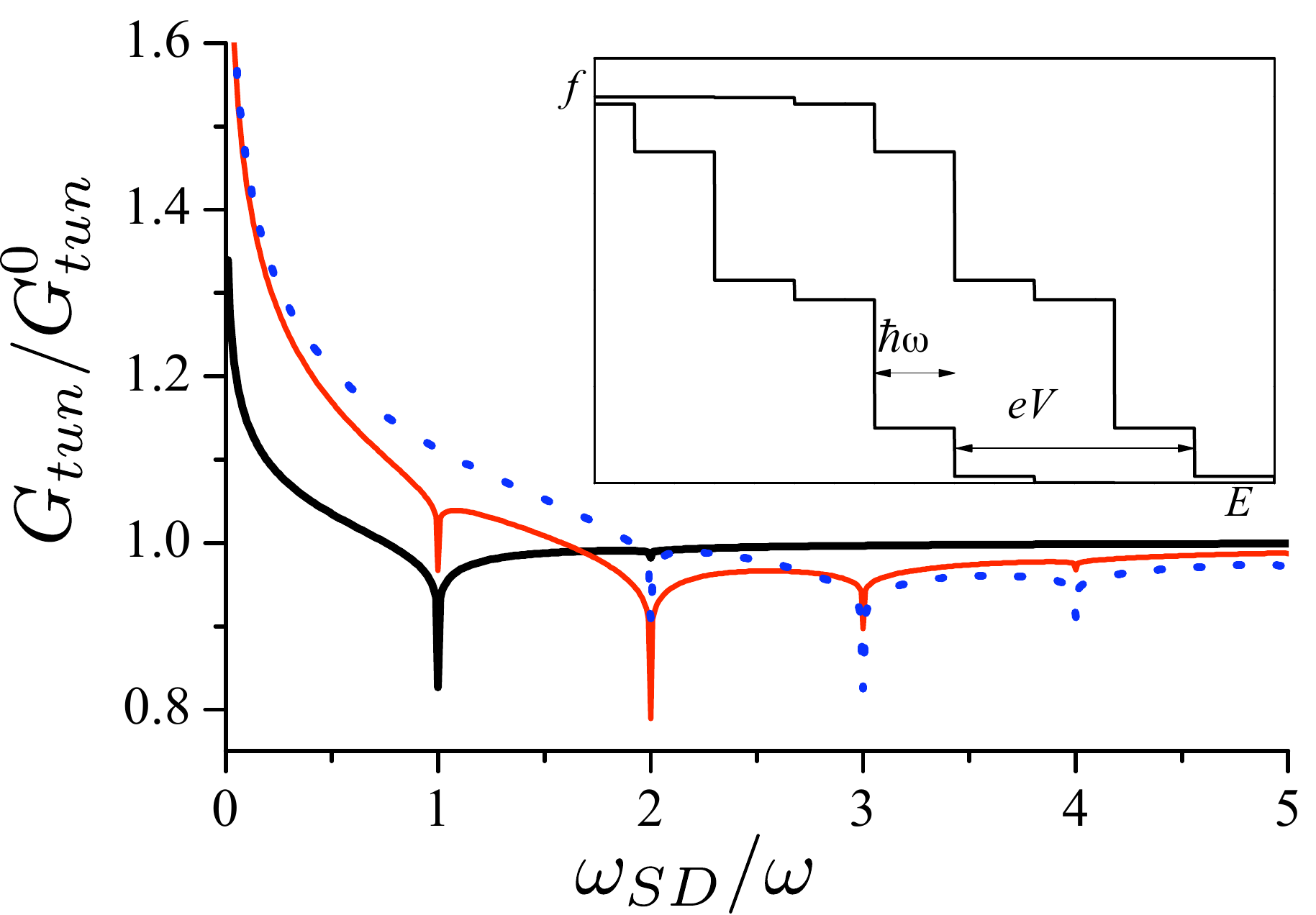} 
\caption{\label{Fig2} (color on-line) The tunneling current across the QPC as a function of the dc bias voltage frequency, normalized to the value $G_{tun}^0=G_{tun}|_{\omega=0}$. The cusps at integer values of the spin gate frequency $\omega$ are a signature of electron-electron interaction (here $K=0.7$). Different curves refer to different values of the spin gate  parameter: $z=1$ (black thick line), $z=3$ (red thin line), and $z=4$ (blue dashed line). Inset: For non-interacting electrons, the two non equilibrium distributions of the electrons incoming to the QPC.  }
\end{figure}

Let us now specify the conditions for this effect to occur. We shall henceforth consider a purely spin gate voltage configuration, i.e. $V_{g,T}=-V_{g,B}=V_{g,s}$, and assume for definiteness  $eV_{g,s}(x,t)=W_{s}(x) A_s(t)$, where $W_s(x)=e V_{g,s}^0$ for $|x|<l_g/2$ and 0 otherwise, and $A_s(t)=\sin(\omega t)$. In terms of the setup in Fig.\ref{fig-setup}, such spin gate voltage corresponds to $V_{g,T}$ and $V_{g,B}$  oscillating with amplitude $V_{g,s}^0$, frequency~$\omega$ and opposite phase. In this case one has $\xi_{s+}^0(t) 
= - (2 \pi)^{-1/2}   4 e V_{g,s}^0  \sin(\omega l_g/4v) \sin(\omega t) /\hbar \omega$. 
We now analyze how such ac gate voltage affects the dc current that flows through the four terminal setup   when electrochemical potentials $\mu_{T/B}^S$ ($\mu_{T/B}^D$) are applied to the Top and Bottom source (drain) metallic electrodes, i.e. on the left (right) hand side of the sample. For simplicity, we focus here on the situation where the electrochemical potential of the electrodes are set in charge-bias configuration, i.e.  $\mu^S_T=\mu_B^S=-\mu_T^D=-\mu_B^D=eV/2$. The current operator  in a Top/Bottom electrode consists of a charge current and a  spin density contribution,
$\hat{I}_{T/B}= \hat{I}^{(c)}\pm \hat{I}^{(s)}$,  given  by $\hat{I}^{(c)}= e  vK  (2 \pi)^{-1/2}  \partial_x\Theta_c$ and 
 $\hat{I}^{(s)}=e  vK  (2 \pi)^{-1/2}   \partial_x \Phi_s$, respectively. The average current $I_{T/B} \doteq \langle \hat{I}_{T/B} \rangle$ can be evaluated with  the Keldysh technique~\cite{keldysh}. To leading order in the tunneling amplitude~$\gamma$, the contribution $I^{(c)}(x,t) \doteq \langle \hat{I}^{(c)}(x,t)\rangle$ at a point $x$ and time $t$ reads\fab{here I have replaced $KV \rightarrow V$}
\begin{eqnarray} 
\lefteqn{I^{(c)}_{\gamma^2}(x,t) =   \displaystyle   - \frac{2 \pi i}{e\hbar}     \left( \frac{\hbar v_F \gamma}{2 \pi a} \right)^2 \int_{t_1 \ge t_2} \hspace{-0.5cm} dt_1 dt_2 \,    \,   \,  \sigma_{0c}(x,0;t_x- t_1)   } \hspace{1.5cm} & & \nonumber \\   
&  &  \hspace{-1cm} \times \left( \sum_{s=\pm} s \, \prod_{\nu=c,s} e^{ 2 \pi   \mathcal{C}_{\nu +}(s(t_1-t_2))   } \right)  \sin \left[ \frac{(t_1-t_2)e V}{\hbar}   \right] \,    \,  \nonumber \\  
&  &   \hspace{-1cm} \times \cos \left[ \sqrt{2 \pi} (\xi^0_{s+}(t_1)   - \xi^0_{s+}(t_2)) \right]    \,  \, \,   
  \label{Ic_gamp2_fin-inf-pre-pre}
\end{eqnarray}
where $\sigma_{0c}(x,y;t)=(2K e^2/h) \, \theta(t)  \sum_{p=\pm}\delta(t +p |x-y|/v)$ represents the charge conductivity,   $\mathcal{C}_{\nu +}(t) \doteq \langle \xi_{\nu +}(0,t) \xi_{\nu +}(0,0)\rangle-\langle \xi^2_{\nu +}(0,0)\rangle =-K_\nu \ln[(t_a+i t)^2/t_a^2]/4\pi$ is the unperturbed correlation function of the charge and spin fields at the tunneling point, and $t_a=a/v_F$ is a small cut-off timescale. The last line of Eq.(\ref{Ic_gamp2_fin-inf-pre-pre}) encodes the phase differences arising from the spin  FS. After obtaining a similar expression  for  $I^{(s)}_{\gamma^2}$,  one can  evaluate the currents, which 
consist of a dc and an ac components, $I_{T/B}(x,t)=I_{dc}+I_{T/B,ac}(x,t)$, where $I_{dc}$ is independent of $x$ and $t$. In particular, $I_{dc}$ can be written as $I_{dc}=I_{0}+I_{\gamma^2}$, where\fab{\bf here} $I_0=e^2 V/h$ is the current in the absence of the QPC, and the $I_{\gamma^2}$ represents the dc  current   tunneling across the constriction,  which depends on the dc bias voltage $V$ and includes the effect of the  ac gate voltage. Computing the tunneling conductance $G_{tun} \doteq dI_{\gamma^2}/dV$  we obtain \fab{(here I have replaced $K^2 \rightarrow K$)}
\begin{eqnarray}  
\lefteqn{\displaystyle G_{tun} =    - \frac{K  e^2}{\hbar^3}   \left( \frac{\hbar v_F \gamma}{2 \pi a } \right)^2  \! t_a^{2K^*}  \sin[\pi (1-2K^*)] }  & & \label{PAT}  \\
& &  \displaystyle \times \Gamma[1-2K^*]     (2K^*-1) \sum_{n \in \mathbb Z} J^2_{|n|}(z) \,   \, | \omega_{SD}-n\omega|^{2(K^*-1)}    
 \nonumber
\end{eqnarray}
where $K^*=(K_c+K_s)/2$ is the effective interaction strength, $J_n$ is the Bessel function,\fab{\bf here} 
$\omega_{SD}=e V/\hbar$ the frequency related to the dc bias voltage, and $z \doteq 4eV_{g,s}^0 \,  \sin( \frac{\omega l_g}{4v} )/\hbar \omega$   a spin gate amplitude parameter. \\Eq.(\ref{PAT}) describes how  the FS processes arising from the ac spin gate voltage affect  the tunneling conductance. Formally, it is reminiscent of expressions obtained in the context of photon-assisted tunneling~\cite{butti}. Here, however,  it is the spin degree of freedom of the tunneling electron to be affected. Notice that  Eq.(\ref{PAT}) is intrinsically gauge invariant, for it only depends on energy differences ($z \propto V_{g,T}-V_{g,B}$). The behavior of $G_{tun}$,  shown in Fig.\ref{Fig2}, consists of a pattern of cusps, located at values of $\omega_{SD}$ integer multiples of the ac spin gate frequency. The weight $J^2_{|n|}(z)$ of each cusp  is a non-monotonous function of the spin gate amplitude $V_{g,s}^0$, and the cusp exponent is interaction  dependent. We observe that, because the spin channel is attractive ($K_s \ge 1$),   the effective interaction parameter is $K^* \ge 1$, and one obtains cusps, and not divergences.
Importantly, in the non-interacting limit, $K^* \rightarrow 1$, the  exponent of the singularities vanishes,  and the Bessel rule $\sum_{n \in \mathbb Z} J^2_{|n|}(z) \equiv 1$ leads $G_{tun} $ to be independent of the spin-gate  parameters $V_{g,s}^0$ and $\omega$.  This can be understood considering the electrons incoming towards the tunneling point.  A time-dependent spin gate voltage locally changes the distributions of the incoming electrons into  strongly non-equilibrium distributions $f_{S/D}(E)  \rightarrow \sum_{n \in \mathbb Z} J^2_{|n|}(z) f(E\mp eV/2-n \hbar \omega)$, where all the harmonics $\hbar \omega$ related to the   gate ac frequency appear [see inset in Fig.\ref{Fig2}]. For a non interacting system the tunneling current is given by $I_{\gamma^2} \sim   \int \! R \, [f_S(E)-f_D(E)] dE$, where $R \propto \gamma^2$ is the energy independent `reflection coefficient' induced by the tunneling term. Thus, although  $f_S$ and $f_D$ are affected by the ac gate voltage, the integral is not. In contrast, when electron-electron interaction is included,  $R$ acquires an effective energy dependence, with a singular behavior at the Fermi energy~\cite{glazman}. The cusps thus emerge whenever the energy difference between the incoming states vanishes, providing a hallmark of electron interaction in QSHE edge states.

We observe that previous works concerning time-dependent impurities~\cite{feldman-gefen,chamon,komnik} have discussed the effect of a time-dependent {\it magnitude} of the BS term, in systems where FS cannot affect dc transport.
In contrast, here we have shown that in QSHE edge states an ac FS term induces a time-dependence in the {\it phase} of the `BS',   even when the magnitude BS term is time-independent. This difference implies important conceptual consequences. First, even for a monocromatic ac gate voltage, all harmonics $n \omega$ appear in the tunneling current, thus broadening to a lower frequency range the possibility to observe resonances with the dc voltage ($\omega=\omega_{SD}/n$). Second, it provides an additional parameter, $V_{g,s}^0$, to modify $G_{tun}$ in a non-monotonous way. 


In conclusion, we have shown that two gate voltages applied at a QPC of a QSHE system lead to two types of FS processes: The first one corresponds to the usual charge gate coupling and has no effect on dc transport, just like the FS term of a single impurity in a quantum wire. The second one, stemming from the helical nature of the QSHE edge states, consists of a spin gate coupling, and does affect dc transport. Indeed an ac spin gate leads to the cusp pattern in the dc tunneling conductance Eq.(\ref{PAT}), shown in Fig.\ref{Fig2}. We have also demonstrated that such pattern is a signature of electron-electron interaction, for it disappears in  the non-interacting case. The obtained tunneling current is reminiscent of the photon-assisted tunneling. In QSHE, however, it is the spin degree of freedom of the tunneling electron to be controlled by the ac gate voltages. This feature, combined with the attractive interaction of the spin sector, represents a unique unconventional effect that is not present in other 1D systems. 
For typical values $l_g \sim 100 {\rm nm}$, $v \sim 5 \cdot 10^{5} {\rm m/s}$, and for $\omega \sim  1{\rm GHz}$, $V_{g,s} \sim {\rm m eV}$ the cusps are located at  bias voltages $eV \sim n \, \mu{\rm eV}$ ($n \in \mathbb Z$),     and the observation of the effect is within reach of  experiments. 
\\We greatly acknowledge fruitful discussions with B.~Trauzettel, C.~Br\"une,   F. Taddei,
and financial support by the VIGONI Program of   Ateneo Italo-Tedesco.


\begin{thebibliography}{77}
\bibitem{nota} Here we limit our study to spin-preserving tunneling; see Refs.\cite{liu,mio} for spin-flip  effects.
\end{thebibliography}
\end{document}